\documentclass{article}

\usepackage{microtype}
\usepackage{graphicx}
\usepackage[labelformat=simple]{subcaption}

\usepackage{booktabs} 
\usepackage[frozencache =true,cachedir=./, newfloat]{minted}

\usepackage[nohyperref, accepted]{icml2022}

\usepackage{amsmath}
\usepackage{amssymb}
\usepackage{mathtools}
\usepackage{amsthm}
\usepackage{xurl}

\usepackage[smallcaps, acronym]{glossaries}
\glsdisablehyper{}

\newacronym{marl}{marl}{multi-agent reinforcement learning}
\newacronym{abm}{abm}{agent based model}
\newacronym{rl}{rl}{reinforcement learning}
\newacronym{ppo}{ppo}{proximal policy optimisation}
\newacronym{ml}{ml}{machine learning}

\usepackage[capitalize,noabbrev]{cleveref}

\SetupFloatingEnvironment{listing}{name=Code Snippet}

\theoremstyle{plain}

\theoremstyle{definition}

\theoremstyle{remark}

\usepackage[textsize=tiny]{todonotes}

\icmltitlerunning{High Performance Simulation for RL}

\begin{document}

\twocolumn[
\icmltitle{High Performance Simulation \\for Scalable Multi-Agent Reinforcement Learning}




\begin{icmlauthorlist}
\icmlauthor{Jordan Langham-Lopez}{comp}
\icmlauthor{Sebastian M. Schmon}{comp}
\icmlauthor{Patrick Cannon}{comp}
\end{icmlauthorlist}

\icmlaffiliation{comp}{Improbable, London, UK}

\icmlcorrespondingauthor{Jordan Langham-Lopez}{jordanlanghamlopez@improbable.io}

\icmlkeywords{Machine Learning, ICML}

\vskip 0.3in
]



\printAffiliationsAndNotice{}  

\begin{abstract}
Multi-agent reinforcement learning experiments and open-source training environments are typically limited in scale, supporting tens or sometimes up to hundreds of interacting agents. In this paper we demonstrate the use of \emph{Vogue}, a high performance \gls{abm} framework. Vogue serves as a multi-agent training environment, supporting thousands to tens of thousands of interacting agents while maintaining high training throughput by running both the environment and \gls{rl} agents on the GPU. 
High performance multi-agent environments at this scale have the potential to enable the learning of robust and flexible policies for use in \glspl{abm} and simulations of complex systems. We demonstrate training performance with two newly developed, large scale multi-agent training environments. Moreover, we show that these environments can train shared \gls{rl} policies on time-scales of minutes and hours.
\end{abstract}

\glsresetall

\section{Introduction}

Increases in computational power and available data have led to an increase in the potential scale and fidelity of computer simulations such as \glspl{abm}. In tandem, a desire has developed for complex agent behaviour which cannot be addressed solely by the often error-prone and challenging process of manually designing and implementing agent behaviours that are both flexible and robust to extreme states of the simulation.

In the context of agent-based models, \gls{marl} is a means of learning agent behaviours where the reward function is well known but the simulation has a complex state space. This motivates the development of fast and large-scale training environments, enabling the use of \gls{marl} to learn behaviours reflecting emergent systems such as crowds, flocks, social networks or markets.

\begin{figure}[t]
\vskip 0.2in
\begin{center}
\centerline{\includegraphics[width=\columnwidth]{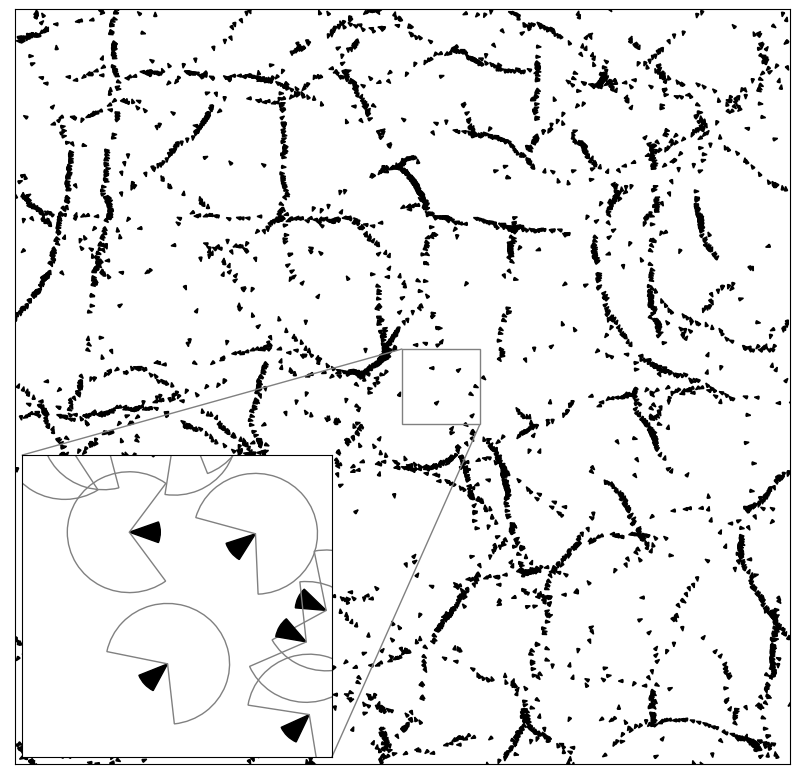}}
\caption{Snapshot of agents in the \emph{flock} \gls{marl} environment. The agents are rewarded for being close to other agents, but are penalised for collisions. This example contains 5,000 agents learning a shared policy. 500 training steps (totalling 625,000 mini batches) took approximately 16 minutes to train on a laptop Nvidia GTX 1650 GPU. The inset illustrates how agents sense neighbouring agents via a simple vision model, shown in more detail in Figure \ref{fig:view_diagram}.
}
\label{fig:tag_scatter}
\end{center}
\vskip -0.2in
\end{figure}

\section{High Performance \gls{marl} with Vogue}
\subsection{Vogue: An Agent-Based Modelling Framework}

Vogue is a commercial Python framework for high performance \glspl{abm}. It is motivated by some key design principles:

\begin{itemize}
    \setlength\itemsep{0mm}
    \item \textbf{Usability:} Vogue allows users to write \gls{abm} logic in native, idiomatic Python, as shown in code snippet \ref{code:boid_update}. This is then compiled by Vogue into high performance bytecode, and can be run in a standard Python process.
    
    \item \textbf{High Performance Execution:} Vogue maps model logic over agents using high performance, vectorised update patterns. This is efficient and obviates the need for modellers to consider the computational complexity and low-level implementation details of \glspl{abm}.
    
    \item \textbf{Composition:} Vogue model logic is described in small pieces of functionality following standard patterns. This allows modellers to easily combine and reuse model logic, and standardises how models are written.
    
    \item \textbf{Interoperability:} Vogue interacts seamlessly with the broader Python data science and \gls{ml} ecosystem. This allows Vogue models to be easily employed in \gls{ml} and data science pipelines. 
\end{itemize}

These design choices allow users to quickly implement high performance \glspl{abm} without the need to write low-level code, as shown in code snippet \ref{code:boid_update}. In comparison to other numerical Python compilation libraries such as JAX \cite{jax2018} or Numba \cite{numba_2015}, Vogue specifically implements common patterns encountered in \glspl{abm}, in particular interactions between spatially embedded agents (based on spatial proximity), and agents interacting via edges on a graph or network. These base interactions can them be combined into more complex model logic, including interactions between heterogeneous agent types.

Vogue is currently limited to models that update in discrete time-steps. It is designed such that individual interactions update agents simultaneously, in particular the GPU engine applies interactions to entities in parallel. 

\begin{listing}[h]
    \begin{minted}[
frame=lines,
framesep=3mm,
baselinestretch=1.2,
% bgcolor=LightGray,
fontsize=\footnotesize
]{Python}

@vogue.interaction.self()
def update_opinion(
    params: Params, person: Person
) -> None:
    person.opinion = person.new_opinion


@vogue.interaction.graph()
def social_influence(
    params: Params,
    me: Person,
    you: Person,
    edge: Friendship,
) -> None:
    d = abs(me.opinion - you.opinion)
    if d < params.threshold:
        w = params.strength * edge.weight
        me.new_opinion = (
            (1.0 - w) * 
            me.new_opinion + 
            w * you.opinion
        )

    \end{minted}
    \vspace{-5mm}
    \caption{A Vogue implementation of a simple opinion dynamics model. The model is implemented as two \emph{interactions}. A self interaction that updates the current opinion of entities, and a graph interaction that updates an entities opinion based on its neighbours opinions. Model logic in Vogue is expressed as interactions between pairs of agents, or updates of individual agents. The compilation of these interactions and execution of the model is handled by the Vogue engine.}
    \label{code:boid_update}
\end{listing}

\subsection{End-to-end GPU-based Reinforcement Learning}

As well as allowing users to efficiently implement models, a major advantage of Vogue is that models can be executed completely on the GPU with minimal changes to the underlying code, leveraging the high degree of parallelism afforded by the GPU to efficiently update the state of the model. Vogue's implementation allows data to be shared between Vogue and other \gls{ml} and deep-learning libraries without copying (or transfer via the host) via the CUDA array interface. This means Vogue can interact directly with any \gls{ml} framework that implements the CUDA array interface, including popular \gls{ml} tools like PyTorch \cite{torch2019} and JAX \cite{jax2018}, or GPU based data tools like RAPIDS \cite{rapids2022}. This allows the execution of both the Vogue training environment/simulation and the deep \gls{rl} agent on the GPU, removing the overhead of moving data to and from the GPU during training. 

\section{Background}

\Gls{marl} has found a wide range of applications including distributed autonomous systems, socioeconomic and game theoretic problems \cite{marl_review_2008}. The number of learning agents in these systems typically ranges from single-digit to hundreds of interacting agents. 

A major obstacle to applying \gls{marl} to systems with larger numbers of interacting agents is the lack of high performance simulation environments, as well as the technical challenges and cost associated with their development. Training environments are often implemented for execution on the CPU with the deep \gls{rl} agent running on the GPU. However, the communication of data from the simulation on the host to the agent on the GPU can have a significant performance impact.


OpenAI Gym \cite{openaigym2016} has become a de facto standard API for \gls{rl} research, allowing training environments and \gls{rl} agent implementations to be easily combined, though it does not have a standard API for multi-agent training. There are a number of open-source \gls{marl} environment projects implementing this API, including Petting Zoo \cite{pettingzoo2020} and MAgent \cite{magent2018}. \mbox{MAgent} potentially supports environments containing millions of agents, but in contrast to Vogue (proposed here) the environments restrict agents to a discrete grid and the engine is implemented in C++. Petting Zoo wraps a number of environments (including MAgent), though many of the environments are restricted to tens or hundreds of agents, or have observation spaces that scale with the number of agents, restricting their usable scale.

RLLib \cite{rllib2017} provides a set of tools for the execution of \gls{rl} pipelines, including automated hyper-parameter tuning and distributed training. It has the ability to run pipelines with multiple multi-agent policies, and multiple environments executed in parallel. It does however rely on users to implement the training environment (or providing a pre-existing environment).

Warpdrive \cite{warpdrive2021} offers a framework for end-to-end \gls{marl} on GPU. It demonstrates high training throughput, though is only tested \citep[in][]{warpdrive2021} on a maximum of 1,000 agents (executed across 2,000 environments simultaneously). Its end-to-end solution contrasts Vogue's focus on ease of writing high performance models, and interoperability with existing Python tools. NVidia's Issac Gym \cite{isaac_2021} takes a similar approach, sharing data between simulation and RL agent on the GPU directly. Their focus is on physics simulations for robotics, where Vogue is focused on \gls{abm} use cases, and ease of implementing and combining \glspl{abm}.

\begin{figure}[t]
\vskip 0.2in
\begin{center}
\centerline{\includegraphics[width=\columnwidth]{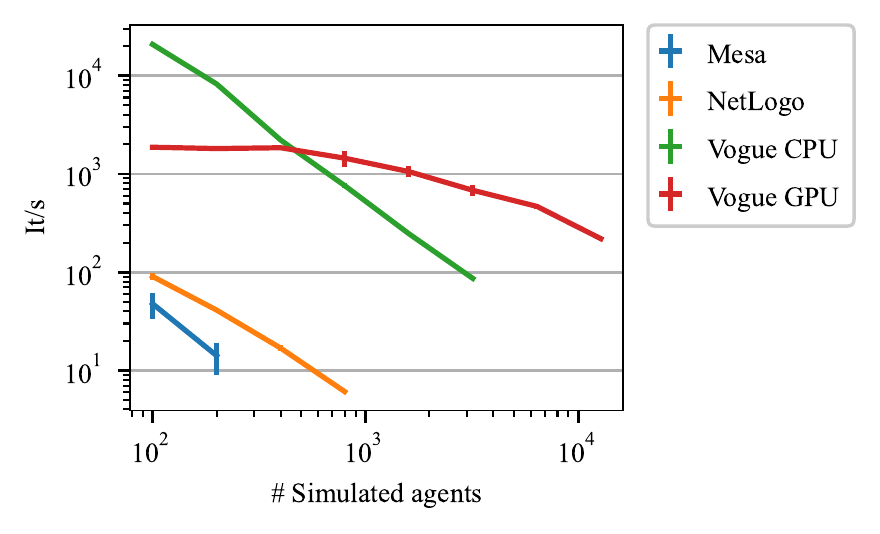}}
\caption{Average updates per second of a simple boids model implemented in each framework. The Vogue GPU engine is initially slower than the CPU but demonstrates much better scaling with increasing number of agents.}
\label{fig:vogue_comparison}
\end{center}
\vskip -0.2in
\end{figure}

There are a number of \gls{abm} frameworks, sharing the broad aim of allowing users to more easily implement \glspl{abm}, or provide performance gains. Mesa \cite{python-mesa-2020} is a popular Python \gls{abm} framework, though its performance is restricted by Python's native performance. Netlogo \cite{netlogo1999} is instead implemented in Java, though this limits its interoperability with Python and requires users to implement models in Netlogo's scripting language. A comparison of the performance of these libraries for a basic implementation of boids (with hard-coded behaviours) is shown in Figure \ref{fig:vogue_comparison}. Both the CPU and GPU version of Vogue are several orders of magnitude faster than Mesa and NetLogo. The GPU engine demonstrates improved scaling over the CPU engine with increasing number of agents. Few frameworks support GPU execution; Flame \cite{flamegpu2021} is a notable example, implemented in C++ with an optional Python interface.

\section{Training Environments}

At the time of writing, we could not locate any documented training environments at the scale of 1,000+ interacting agents. Instead, we provide two simple examples by implementing extended versions of two popular models: \emph{flocking} and \emph{tag}. We note that extending the range and complexity of example environments is a priority for future work.

\begin{figure}[ht]
\vskip 0.2in
\begin{center}
\centering
    \begin{subfigure}[b]{0.55\columnwidth}
        \includegraphics[width=\textwidth]{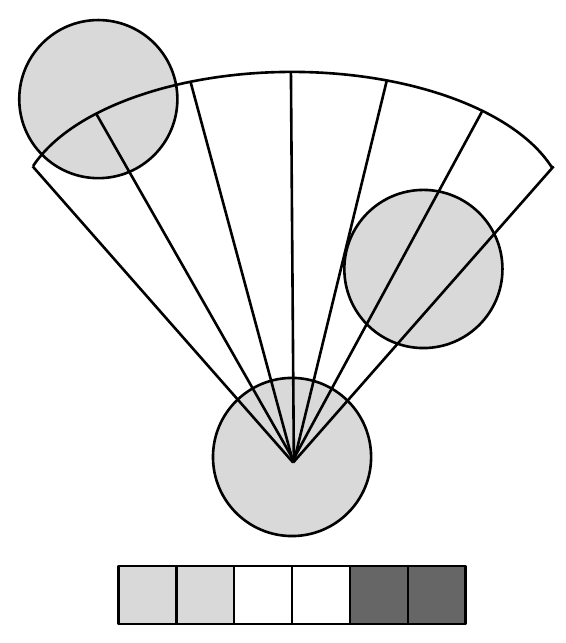}
        \caption{Agent vision model}
        \label{fig:view_diagram}
    \end{subfigure}%
    \hfill
    \begin{subfigure}[b]{0.4\columnwidth}
        \includegraphics[width=\textwidth]{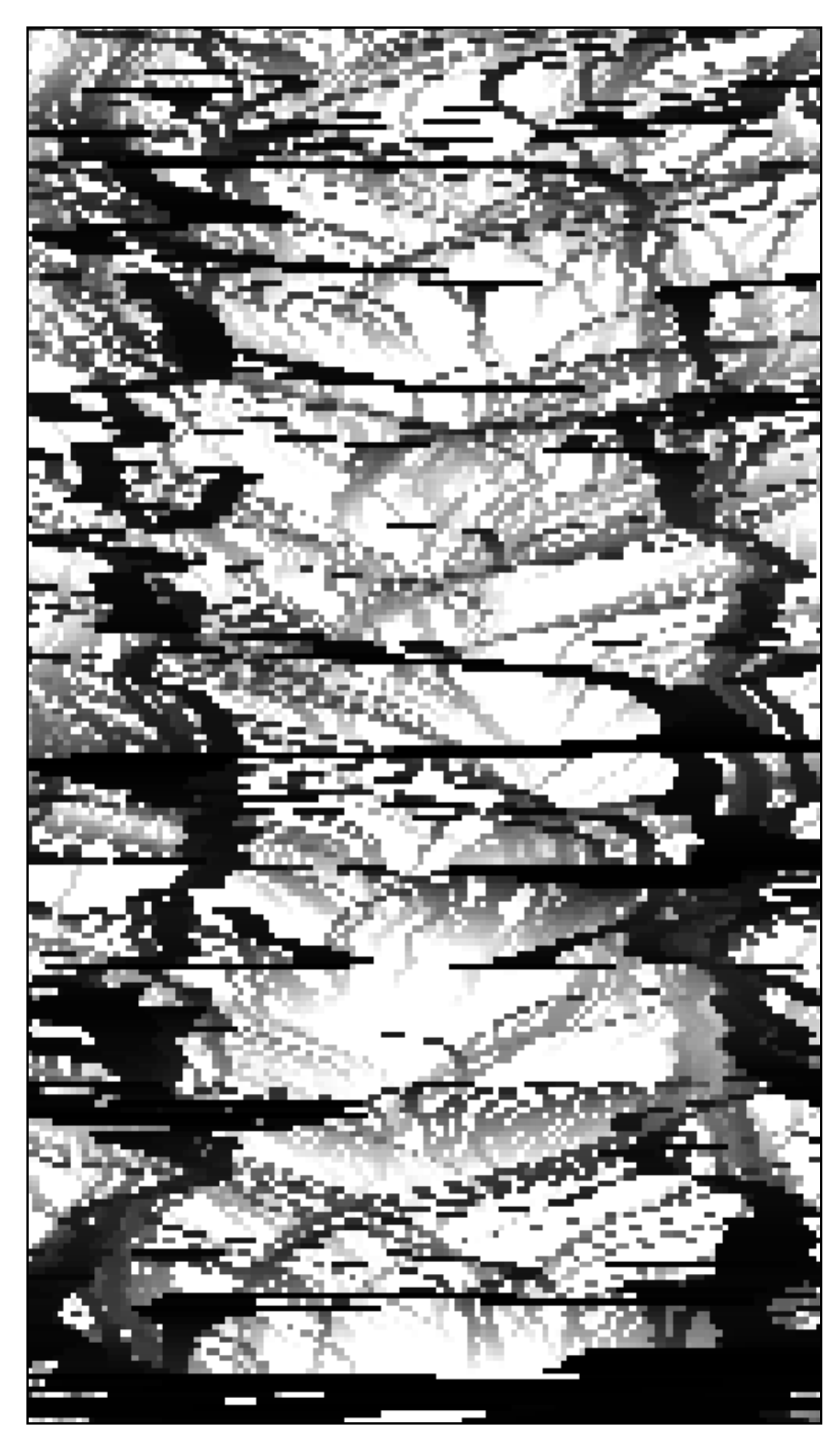}
        \caption{Agent's view over time}
        \label{fig:boid_fov}
    \end{subfigure}%
\caption{
Agents visualise their local environment using a simple ray-casting model, segmenting their field of vision into a fixed number of cells \subref{fig:view_diagram}. The resulting array (shown below the agent) represents the distance to the nearest neighbouring agent in the simulation. An example time series of an agent's view is shown in image \subref{fig:boid_fov}; darker areas are objects closer to agent.
}
\label{fig:agent_view}
\end{center}
\end{figure}

In both examples, to address the increased overall size of the simulation state space, the agents only observe their local neighbourhood of the simulation via a simple ray casting algorithm that divides an agent's field of view into a fixed number, $v$, of sectors. Internally, their view is represented by an array of $v$ values representing the distance to the nearest object at that angle, as shown in Figure \ref{fig:view_diagram}. An example time series showing an agent's view during a training step is shown in Figure \ref{fig:boid_fov}. This local view forms part of the training environment observation space, local to each agent in the environment.

\subsection{Flocking}

This training environment is based on Reynolds' Boids model \cite{boids1987}, designed to mimic natural flocks and shoals.
The training environment consists of a flock of agents navigating a 2d space, with no hard-coded navigation rules. Previous studies have looked at emergence of flocking from reinforcement learning \citep[for example][]{durve2020, shimada_2018} but in this work we aim to examine the feasibility of efficiently training flocking behaviours for 1,000-10,000+ interacting agents. A snapshot of a simulation state is shown in Figure \ref{fig:tag_scatter}.

\begin{figure}[ht]
\vskip 0.2in
\begin{center}
\centerline{\includegraphics[width=\columnwidth]{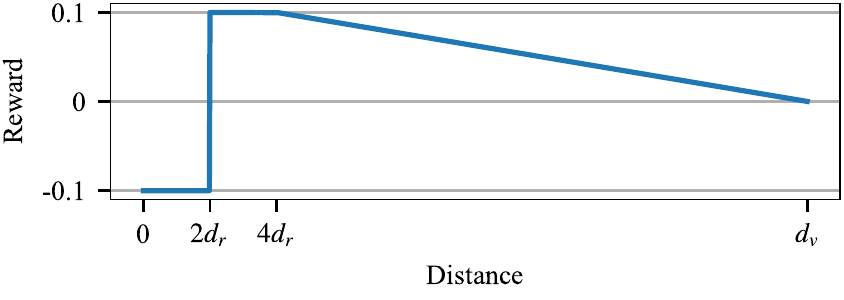}}
\caption{Flock reward function, $f$, where $d_{r}$ is the agents' radius, and $d_{v}$ the agents' vision range. Agents are penalised when they collide at a distance of $2d_{r}$}
\label{fig:flock_reward}
\end{center}
\vskip -0.2in
\end{figure}

The observation space of the environment (for an individual agent) is a 129 length vector containing the agent's view, with $v=128$ sectors, and the agent's current speed. The action space is two dimensional. One dimension accelerates/decelerates the agent, in the range $[-a_{\text{max}}, a_{\text{max}}]$, updating the speed of the agent as $s_{t+1}=\text{min}(s_{\text{min}}, \text{max}(s_{t} + a_{t}, s_{\text{max}}))$. The second dimension in the range $[-\theta_{\text{max}}, \theta_{\text{max}}]$ rotates the heading of the agents up to a maximum rotation rate. The reward signal for each agent is the sum of contributions due to proximity to other agents, i.e. the reward $r_{i}$ of agent $i$ is

\begin{equation}
    r_{i} = \sum_{j}f(d_{ij})\quad\text{for}\quad
    d_{ij} < d_{v},
\end{equation}

where $d_{ij}$ is the Euclidean distance between agents $i, j$, $d_{v}$ is the visual range of the agent, and $f$ is the reward contribution, as in Figure \ref{fig:flock_reward}.

In each environment update, the agents are accelerated and rotated (according to actions sampled from the current policy), after which their local view model and rewards are updated for all pairs of agents in spatial proximity.

\subsection{Tag}

This environment is based on Petting Zoo's \citep[in][]{pettingzoo2020} \emph{simple tag} environment. It contains two agent types: \emph{runners} and \emph{chasers}. Chasers are rewarded for touching runners, and runners are rewarded for staying close to other runners, and penalised for touching a chaser. The view model of the agent has two simulated colour channels allowing agents to distinguish between neighbouring runners or chasers. The corresponding observation space is a $128$ length vector, concatenating the two $v=64$ colour channels. The two dimensional action space rotates the agent's heading, in the range $[-\theta_{\text{max}}, \theta_{\text{max}}]$ and moves them along the heading, in the range $[0, s_{\text{max}}]$ where $s_{\text{max}}$ is the maximum speed of the agents.

\subsection{Training}

\begin{figure}[th]
\vskip 0.2in
\begin{center}
\centerline{\includegraphics[width=0.6\columnwidth]{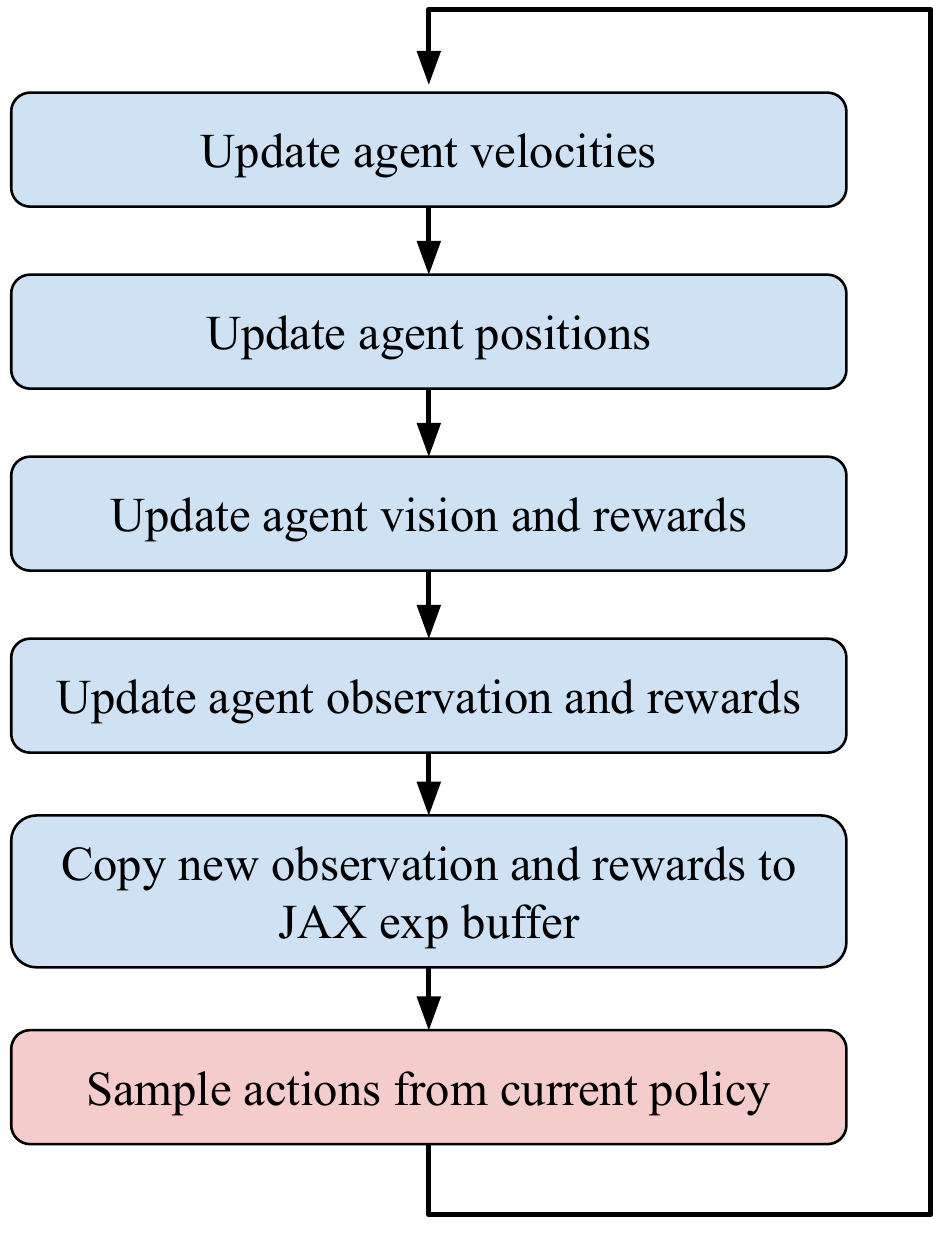}}
\caption{The experience collection loop. Vogue (blue) updates the state of the \gls{abm} in response to the actions sampled from the PPO policy implemented in JAX (red). It then updates each agent's local view and rewards then copies them to the JAX experience buffer.}
\label{fig:rl_loop}
\end{center}
\vskip -0.2in
\end{figure}

For both environments we used \gls{ppo} \cite{ppo2017} for a continuous action space, with the simulated agents sharing a single policy. In the tag environment, the runner and chaser types share independent policies, trained simultaneously. Instead of multiple parallel environments, we treat each individual agent as an independent learner, and gather trajectories across the full set of of agents. If we have $n$ agents and run $t$ updates of the simulation each training step, we sample a total of $n\times t$ trajectories per training step. Vogue updates the state of the \gls{abm} in response to the actions sampled from the PPO policy, then copies the updated observations and rewards to the JAX experience buffer, as shown in Figure \ref{fig:rl_loop}.

\section{Performance Results}

Benchmarking was performed on an Nvidia A100 GPU. Each full training run includes $T=200$ training steps. Inside each training step the environment was updated for $t=128$ steps, generating $m = n\times t$ individual trajectories. The policy was then updated for $p=2$ epochs, across $b$ mini-batches sampled from the trajectories. The number of trajectories scales with the number of simulated agents, so the number of mini-batches is capped as $b=\text{min}(m/|b|, b_{max})$ where $b_{max}=512$ is the maximum number of batches, and $|b|=512$ is the chosen mini-batch size. This means for the highest number of entities each training run performs $T\times p\times b = 204,800$ mini-batch updates of the \gls{ppo} policy, and $T \times t = 25,600$ updates of the environment. The agents have a view distance equivalent to 1/10th of the width of the environment and a convex field of vision of approximately $250^{\circ}$. The \gls{rl} agent update, and trajectory processing algorithms were implemented in JAX. The actor and critic \gls{ppo} networks had two hidden layers with 64 nodes each, with a total of 1035 trainable parameters.

It should also be noted that performance of the simulation can be highly dependent on the parameters of the simulation. For example the view distance of the agents and their vision angle impact the number of interactions between agents that must be processed. 

\begin{figure}[ht]
\vskip 0.2in
\begin{center}
\centerline{\includegraphics[width=\columnwidth]{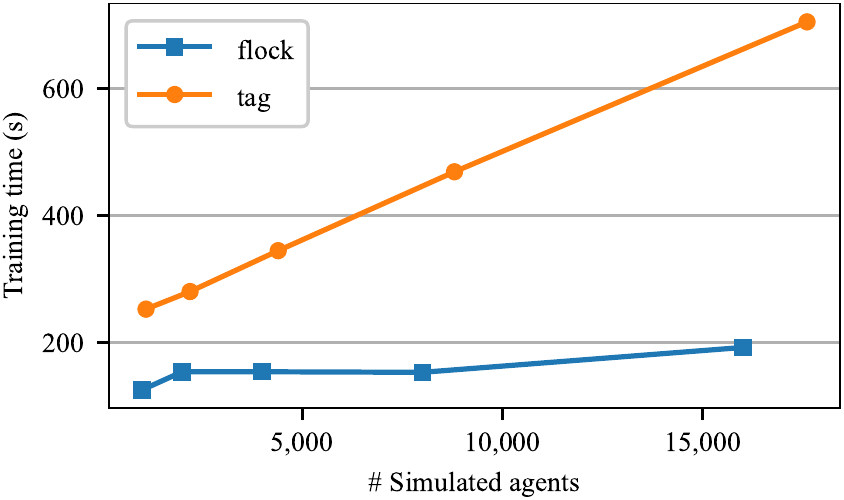}}
\caption{Total training time for 200 training steps, for the flock and tag environments, with increasing number of agents. Each training step includes running the model for 128 steps with the current policy, and updating the policy over 1024 mini-batches. Results are shown averaged over 10 training runs, though error bars are omitted due to their small relative size. For the tag environment 1/10th of the shown number of agents are chasers.}
\label{fig:total_train_time}
\end{center}
\vskip -0.2in
\end{figure}

The total training time for 200 steps using the flock and tag environments are shown in Figure \ref{fig:total_train_time}. Values shown are averaged over 10 independent training runs. With 16,000 agents, the flock environment takes around 3.5 minutes to complete training, and demonstrates slow scaling with number of agents. For a tag environment with a total of 17,600 agents (16,000 runners and 1,600 chasers) training takes around 12 minutes. The tag environment demonstrates roughly linear growth in training time with number of agents, we speculate because of the added complexity of multiple interactions between different agent types. 

\section{Conclusion}

We have demonstrated that joining forward simulation and training of \gls{marl} systems into a single end-to-end \gls{rl} process can lead to drastic improvements in performance. Using the commercial software \emph{Vogue}, high performance \glspl{abm} can be executed directly on the GPU, while the high level of interoperability enables direct communication between Vogue and many Python based \gls{rl} or deep learning frameworks.
Vogue is currently available for academic research -- please contact the authors.

\bibliography{references}
\bibliographystyle{icml2022}
\end{document}